\documentclass[fleqn,10pt]{wlscirep}
\usepackage{amsmath,amssymb}
\usepackage{hyperref}
\usepackage{bbold}
\usepackage{float}
\usepackage{hyperref}
\usepackage[utf8]{inputenc}

\title{Multivariate analysis of short time series in terms of ensembles of correlation matrices}

\author[1,*]{Manan Vyas}
\author[2,3,**]{T. Guhr}
\author[1,3,***]{T. H. Seligman}
\affil[1]{Instituto de Ciencias F{\'i}sicas, 
Universidad Nacional Aut{\'o}noma de M\'{e}xico, 
62210 Cuernavaca, M\'{e}xico}
\affil[2]{Fakult\"at f\"ur Physik, Universit\"at Duisburg-Essen, Lotharstra$\beta$e 1, D-47048, Duisburg, Germany}
\affil[3]{Centro Internacional de Ciencias, 62210 Cuernavaca, M\'{e}xico}
\affil[*]{manan@icf.unam.mx}
\affil[**]{thomas.guhr@uni-due.de}
\affil[***]{seligman@icf.unam.mx}

\begin{abstract}
When dealing with non-stationary systems, for which many time series are available, it is common to divide time in epochs, i.e. smaller time intervals and deal with short time series in the hope to have some form of approximate stationarity on that time scale. We can then study time evolution by looking at properties as a function of the epochs. This leads to singular correlation matrices and thus poor statistics. In the present paper, we propose an ensemble technique to deal with a large set of short time series without any consideration of non-stationarity. Given a singular data matrix, we randomly select subsets of time series and thus create an ensemble of non-singular correlation matrices. As the selection possibilities are binomially large, we will obtain good statistics for eigenvalues of correlation matrices, which are typically not independent. Once we defined the ensemble, we analyze its behavior for constant and block-diagonal correlations and compare numerics with analytic results for the corresponding correlated Wishart ensembles. We discuss differences resulting from spurious correlations due to repetitive use of time-series. The usefulness of this technique should extend beyond the stationary case if, on the time scale of the epochs, we have quasi-stationarity at least for most epochs.
\end{abstract}
\begin{document}

\flushbottom
\maketitle

\thispagestyle{empty}

\section*{Introduction}

Non-equilibrium stationary states (NESS) have attracted large amounts of attention in recent years \cite{Ka-84, De-07, Pr-11, Li-04, St-16, Bi-17} but more recently increasing attention is given to non-stationary situations, as they actually cover a wide range of observational as well as of experimental data. Such data cover diverse fields including astronomy, financial markets and meteorology or chemical engineering, fractures and colloids, as well as numerical results for models of such systems and dynamical systems and many others. Among such systems the ones that have several near stationary states with more or less abrupt transitions are of particular interest. Such systems are wide spread and of relevance. They include bi-stable, and multi-stable systems with smooth transitions as well as systems that might run into catastrophic instability. We can think of both types occurring as first order phase transitions under temperature change depending on conditions. Beyond that, we may hope that non-stationary systems may be quasi-stationary over sufficiently short time periods. Yet abrupt non-stationarities may occur and we may hope to obtain either warnings or at least post-event learning from a correlation analysis of known facts over a short time period before the abrupt events.

For the sake of illustration, let us think of a chemical reactor that should produce certain end products in a stationary fashion, but in fact the state is only quasi-stationary. This reactor may have other states that produce less of the desired and more undesirable products and a transition might prove costly. Yet this might get much worse if breaking stationarity may lead to explosions with release of toxic substances, that may in addition cause great cost of lives and health, such as in Bhopal 1984 \cite{Br-05}. To use a Wishart model as a model for non-stationarity was first put forward in \cite{Sc-13} and used for credit risk analysis in \cite{Sc-14, Sc-15}.

Our interest was triggered by studies of financial markets, where the very attempt to define states of quasi-stationary evolution is relatively new \cite{Mu-12}. In this paper, the correlation matrix of short time series was detected to be a good basis to specify the states and clustering techniques were used to identify these states. An attempt to detect conditions under which change may occur was not made and may also be futile in this context, as the clustering technique by definition assigned each correlation matrix to a state and thus borders become unclear. One could use different clustering algorithms to detect larger differences in clusterings, but this might depend very much on the definition of distance we use \cite{book}. 

Time series analysis is rather straight forward for stationary systems, even if these are out of equilibrium (NESS). It can also be extended to analyze non-stationary states using standard detrending techniques to eliminate log time trends and periodic oscillations. These tasks are more challenging if we have many time series.  Nonetheless, it becomes much more difficult to identify state changes \cite{Mu-12} or have some early signals of catastrophic events \cite{arxiv-18}. The most usual way out in complicated non-stationary situations is to assume stationarity over short time intervals. Early attempts in this direction analyze the entire correlation matrix.  This matrix is not invariant and thus the ordering, i.e. the labeling of the time series, becomes very relevant. While for financial markets there is some ordering that has long standing merits in other cases (e.g. for two dimensional array of detectors) this is no longer the case. If we detect different properties, say pressure and temperature, it is entirely unclear if we should give preference to the spatial distribution or to the different types of measurements to assigned indices in the matrix. The use of sophisticated data handling tools will not remove the arbitrariness of basis dependence. For these reasons (and probably because of heritage from the physics background of the authors), it seems reasonable to choose invariant quantities (not only under permutations but also under orthogonal or unitary transformations) that correspond to linear changes in the measuring devices. The logical choices are eigenvalues and eigenvectors. Obviously, the eigenvectors are basis dependent but they may provide relevant information if the preferred basis is reasonable.    

We believe that in some sense eigenvalues do indicate very relevant aspects of dynamics and recently it was shown, that this is also true for the correlation matrices. Using Metropolis dynamics the larger eigenvalues of the correlation matrix of a 2-D Ising model at critical temperature, display a power law, that can be directly derived from the power law of space correlations in this system \cite{Pr-14}; it was further shown  that such a power law will survive if a sufficiently large random subset of time series is used. Yet long time series are essential to see this effect because the number of large eigenvalues rapidly becomes too small as the correlation matrix becomes more and more singular with shorter time series. 

We define a short time correlation matrix as the one for which the time horizon $T$ (length of the time series) is much smaller than the number $N$ of time series. This obeys to the idea that we can have information on a complex dynamical system if we measure more properties. By definition, the correlation matrix will have only $T-1$ (except for $T=1$) non-zero eigenvalues. Thus, increasing $N$ but not $T$ will not affect the number of non-zero eigenvalues. It is important to mention that, given a $N \times T$ data matrix for $N$ time series of length $T$ with $N >> T$, one can always diagonalize the $T \times T$ dimensional correlation matrix of the position series $A^t A/N$ rather than the $N \times N$ correlation matrix of the time series $A A^t/T$ as they have same non-zero eigenvalues. However, as we obtain only $T$ non-zero eigenvalues, we do not obtain good statistics and the analysis is inconclusive - whether the spectral features are noise based or have correlations. Importantly, the correlation matrix of the position series $A^t A/N$ is capable of detecting lead-lag and other non-Markovian effects \cite{Si-04, Si-06, Bu-05, Wa-15}. One can also increase both $N$ and $T$ keeping $N/T$ fixed. While this limit is nice for theoretical purposes, financial markets, chemical reactions, neuron processes etc. have time scales which determine $T$. Thus, we have to either return to the idea of treating the entire correlation matrix or to use some technique to obtain a large enough set of eigenvalues to get good statistics. One option is to use the power map technique. The use of the power map originally introduced for noise reduction \cite{Gu-02, Sc-10} was suggested in \cite{Vi-13} as an appropriate tool to detect correlations if powers very near to identity are used, indeed in \cite{Pr-14} this can be explicitly seen. Yet while the power map does detect correlations efficiently, a detailed understanding of the correlations via the additional emerging spectrum is not yet available. Importantly, information as to the nature of the correlation at this point has not been given and the power map is not transparent due to its  nonlinearity.

The method proposed in the present paper is a more natural one. We shall replace the large but singular $(N >> T)$ correlation matrix by an ensemble of smaller non-singular correlation matrices. This is achieved by first defining an ensemble of data matrices, i.e. sets of $m( << N)$ time series, chosen from a large number of random selections among the given $N$ time series (corresponding to the large but singular correlation matrix). While making selections, we ensure that no time series is repeated in a given data matrix and no two data matrices are the same in the ensemble. There are non-trivial dependencies in the resulting matrices due to random choices. Using this ensemble of the data matrices, we construct the corresponding ensemble of non-singular correlation matrices. In principle, the number of members in the ensemble is given by $\binom{N}{m}$ and thus, the number of non-zero eigenvalues we have can now be increased dramatically. Thus, the method provides higher statistical significance by retrieving information from the given singular data matrix. In this way, we explore the entire position space and thus, obtain the distribution for each of the eigenvalues rather than a single number. While probably we don’t retrieve the full information of eigenvectors we thus reach simultaneously two goals - we obtain smooth distributions and obtain the information about the distribution of each of the eigenvalues. This is particularly relevant for outliers. It implicitly gives information about the outliers among the time series, as these will only appear in some of the subsets. We shall focus on the largest eigenvalues in our examples. Keeping in mind that finally there is no more information available than there is in the original data matrix, we arbitrarily choose the size of the ensemble in order to produce good statistics i.e. we smooth the curves. One can also construct an ensemble of completely independent data matrices in the very large $N$ limit to obtain the ensemble of non-singular correlation matrices and we mention this limit on occasions. It avoids some deviations due to the dependencies but reduces the smoothing. 

In the next section, we describe in detail the construction of the ensemble. In the following section, we will present basic results obtained from supersymmetric calculations to derive the formula for correlated Wishart ensembles with arbitrary correlations. We treat in some detail the case of constant correlation Wishart matrices including the zero correlation case that provides the unbiased a priori hypothesis to which experimental data can be compared to obtain clarification of the data. In situations where average correlations are important, one can also use correlated matrices as a priori hypothesis. We compare numerics with the result obtained from supersymmetric calculations. Here, we will also discuss the differences between the proposed random choice selections resulting in dependencies and completely independent choice for which plenty of analytical results are known. As block structures are important, at least in econophysics, we analyze the special case of block-wise correlated subsets of time series. We compare the supersymmetric results with numerical calculations where we restrict the block situation to two blocks of different size and block-wise constant correlations. We see that the bulk of the spectra is well-described by the analytics while the outliers will only be approximated as far as their average position is concerned. The shape is different as the analytic result we present is for independent time series with large $N, \; T$ and a fixed ratio $\kappa = N/T$. Finally, we give conclusions and an outlook. 

\section{Construction of the Ensemble}

For short time series, the correlation matrices will be strongly singular i.e. the number of non-zero eigenvalues will be greatly reduced and the eigenfunctions corresponding to zero eigen sub-space are arbitrary. We will now introduce a method to overcome these shortcomings.

The building blocks for the correlation(covariance) matrices are rectangular $N \times T$ data matrices $A = [A_{ij}]$, with $i = 1, 2, \ldots, N$ and $j = 1, 2, \ldots, T$. Each row in the data matrix $A$ is a time series of length $T$, measured at usually equidistant times. It can be obtained from observations or experimental measurements of observables like stock prices, temperature, intensity, astronomical observations and so on. The matrix $C = A A^t/T$, with $A^t$ denoting the transpose of matrix $A$, is the $N \times N$ covariance matrix. Wishart matrices are random matrix models used to describe universal features of covariance matrices \cite{Wi-28}. We consider the case for real entries $A_{ij}\in \mathbb{R}$, known in the literature as Wishart orthogonal ensemble (WOE).  For WOE, the matrix elements of $A$ are real independent Gaussian variables with fixed mean $\mu$ and variance $\sigma^2$ i.e. $A_{ij} \in {\mathrm N}_{\mathbb{R}}(\mu,\sigma^2)$. In order to arrive at correlation matrices, one needs to normalize $\mu = 0$ and $\sigma^2 = 1$. In the context of time series, $C$ may be interpreted as the correlation matrix, calculated over stochastic time series of time horizon $T$ for $N$ statistically independent variables. By construction, $C$ is a real symmetric positive semidefinite matrix. For $T < N$, $C$ is singular and has exactly $(N-T-1)$ zero eigenvalues. Note that, stationarity improves when short time series are used. In real applications, one needs to understand the role of correlations and thus, correlated WOE (CWOE) models provide the null hypothesis. CWOE is defined by real-symmetric matrices ${\cal C} = {\cal A} {\cal A}^t/T$, with ${\cal A} = \chi^{1/2} A$. Here, $\chi$ is a real symmetric positive definite non-random $N \times N$ matrix that accounts for the correlations in time series (rows) of data matrix ${\cal A}$ and $A_{ij} \in {\mathrm N}_{\mathbb{R}}(0,1)$. On ensemble average, ${\overline{\cal C}} = \chi$. 

We analyze highly singular correlation matrices ($N >> T$) by constructing ensembles of correlation matrices from a given correlation matrix by randomly selecting short observational time series. By randomly choosing $m$ rows out of $N$ given rows of $A$(${\cal A}$) such that $m = a T$ with $a$ being a real number close but smaller than unity, we construct an ensemble of $m \times T$ dimensional matrices. While making selections, we ensure that no two rows are same in a given matrix and no two matrices are same in the ensemble. Using these, we obtain an ensemble of $m \times m$ non-singular correlation matrices and analyze eigenvalue distribution. 

If the number $N$ of time series available is large compared to the number of entries $T$ in each time series, the discussion of eigenvalues becomes statistically unsatisfactory. A typical example would be financial time series of increments of 40 consecutive closing prices for a selection of $N=400$ shares from some index (say a selection from Standard and Poors 500). In this case we would obtain but 39 non-zero eigenvalues (40 for covariance matrices) from the $400 \times 400$ correlation matrix, which might be all over the place. We propose to select $m$ time series (experimental, observational or computational) with $m < T$ at random. If we allow all different choices, we would end up with a very large ensemble of correlation matrices (in our example, we might choose $m=36$ leading to $\binom{400}{36} \approx 2.5 \times 10^{51}$ choices, which is an unpractically large number). So we choose a random subset of a few thousand and get excellent statistics for eigenvalues. Having more members in the ensemble would increase the amount of spurious information, which enters unavoidably if we allow repeated time series in different members of the ensemble. If on the other hand we do not allow repetitions the results would depend very much on the selection we make and statistics would be less adequate. An alternative may be to make an ensemble of ensembles with different but totally independent choices, and calculate averages and variances of specific statistical quantities obtained for the lower level ensemble. We choose not to go this more complicated route. 

The question arises, how stable and informative the corresponding results are. The purpose of the present paper is to take this simple idea and compare it to cases where analytic results can be derived from well-known results \cite{PRL-10, JSP-12}. We start by analyzing white noise time series and the resulting correlation matrices known as the Wishart ensemble  \cite{Wi-28, Mehta} as well as for correlated Wishart ensembles with constant correlations \cite{Vi-10}. Here, the level densities are known analytically and the $n$-point correlation function converges to the universal result \cite{Mehta}. Because the case of constant correlations will mimic real situations only very roughly, we shall study in more detail the situations where subsets of time series are more correlated among each other than with the time series of other subsets. This will be the typical case of market sectors of stock exchanges. To emphasize the characteristics of such a block structure, we shall restrict ourselves in graphical displays to two blocks in this paper.

We shall see that clear signatures of the correlations (or lack thereof) can be obtained with very good statistics. This distinguishes the present linear method both from the clustering techniques \cite{Mu-12} and the power-map technique \cite{Vi-13}, which are inherently non-linear. The first is a transparent standard technique but requires considerable previous insight into the problem on hand, while the second turns out to be quite stable but interpretation is an open problem. 

\section{Supersymmetry approach}

Time series analysis is an imperative tool to study dynamics of variety of complex systems. Wishart correlation matrices are standard models employed for statistical analysis of ensembles of time series. We provide here a brief sketch of the derivation using standard supersymmetric steps; for further details refer to \cite{Sup1, Sup2, PRL-10, JSP-12}.

In multivariate analysis, it is desirable to derive a ``null hypothesis'' from a statistical ensemble to understand the measured eigenvalue density of the given correlation matrix. The random matrix ensemble we consider is CWOE with arbitrary correlations that gives the 'empirical' (population) correlation matrix $C_0$ upon averaging over the probability density function $P(A|C_0)$ (normalized to unity),
\begin{equation}
P(A|C_0) = [2\pi \mbox{det}(C_0)]^{-T/2} \exp\left[- \displaystyle\frac{1}{2} \mbox{tr} (A^t C_0^{-1} A) \right] \;.
\label{eq-1}
\end{equation}
By construction, $\int d[A] P(A|C_0) AA^t/T = C_0$, with measure $d[A] = \prod_{i=1}^N \prod_{j=1}^T dA_{ij}$ being product of differentials of all independent elements in $A$. It is important to mention that, in the supersymmetric approach, one assumes $T \geq N$ to ensure invertibility of $C_0$. In order to be able to derive the ensemble averaged eigenvalue density (one-point function), we may replace $C_0$ by diagonal matrix $\Lambda$ of its eigenvalues $\left\{\Lambda_1, \Lambda_2, \ldots, \Lambda_N\right\}$ since the domain ${\mathbb{R}}_{N \times T}$ of $A$ is orthogonally invariant.

In terms of resolvent, the ensemble averaged eigenvalue density for correlation matrix $AA^t$ is defined by
\begin{equation}
S(x) = \displaystyle\frac{1}{N\pi}\displaystyle\lim_{\epsilon \to 0+}  \Im \left[\int d[A] P(A|\Lambda) \mbox{tr} \left( \displaystyle\frac{\mathbb{1}_N}{(x-i\epsilon)\mathbb{1}_N - AA^t}\right) \right] \;.
\label{eq-3}
\end{equation}
In case of CWOE (also WOE), the eigenvalue density for the correlation matrices is derived using the supersymmetry technique \cite{PRL-10, JSP-12}. In this approach, the eigenvalue density is written as the derivative of the generating function. The generating function in turn is mapped onto a suitable superspace which leads to drastic reduction in degrees of freedom. Then, the eigenvalue density is derived by introducing eigenvalue coordinates for the supermatrix and integrating over the anti-commuting Grassmann variables.

The generating function $Z$ as a function of source variable $J$ is the starting point of this  approach,
\begin{equation}
Z(J) = \int d[A] P(A|\Lambda) \displaystyle\frac{\mbox{det}(x^+ \mathbb{1}_N + J \mathbb{1}_N- AA^t)}{\mbox{det}(x^+ \mathbb{1}_N - AA^t)}\;; \;\; x^+ \in \mathbb{C} \;.
\label{eq-4}
\end{equation}
Note that $x^+=x+i\epsilon$ and $Z(0)=1$. The one-point function is then computed by the derivative,
\begin{equation}
S(x) = -\displaystyle\frac{1}{\pi  N} \left.\displaystyle\frac{\partial Z(J)}{\partial J}\right|_{J=0} \;.
\label{eq-5}
\end{equation}
The generalized Hubbard-Stratanovich transformation \cite{HS, HS2} and superbosonization formula \cite{SB} have been used to express the generating function as an integral over a suitable superspace. In fact, these are equivalent \cite{EQ}. The determinant in the denominator of Equation \eqref{eq-4} can be expressed as a Gaussian integral over a vector in ordinary commuting variables. Similarly, the determinant in the numerator can be expressed as a Gaussian integral over a vector in anti-communting variables. Combining these expressions, we obtain a Gaussian integral over a rectangular supermatrix $\mathcal{B}$ which is $n \times (2|2)$ dimensional,
\begin{equation}
\mathcal{B} = [u_a,v_a,\zeta^*_a,\zeta_a], \;\;\;\;\;\;\mathcal{B}^t = \begin{bmatrix} u_b \\ v_b \\ \zeta_b \\ -\zeta_b^* \end{bmatrix}_{1\leq b \leq n}\;.
\label{eq-7}
\end{equation}
Here, $\zeta_i, \zeta^*_i$ are Grassmann variables. Using this and $d[\mathcal{B}]=(2\pi)^{-N} \sum_{i=1}^N du_i dv_i \partial\zeta^*_i \partial\zeta_i$, $u_i,v_i \in {\mathbb{R}}$ in Equation \eqref{eq-4} and performing the Gaussian integral over $A$, we apply the duality relation between ordinary spaces and superspaces $\mbox{det}(\mathbb{1}_N+i\mathcal{B}\mathcal{B}^t\Lambda)=\mbox{sdet} (\mathbb{1}_4+i\mathcal{B}^t\Lambda\mathcal{B})$, one can then rewrite the determinant as a superdeterminant. Importantly, the supermatrix $\mathcal{B}^t\Lambda\mathcal{B}$ is $4\times4$ dimensional and the original matrix $\mathcal{B}\mathcal{B}^t$ is $N \times N$ dimensional. This dimensional reduction is the advantage of the supersymmetry technique. The left upper block (boson-boson block) of supermatrix $\mathcal{B}^t\Lambda\mathcal{B}$ is a Hermitian matrix. We now use the generalized Hubbard-Stratonovich transformation to replace the supermatrix $\mathcal{B}^t\Lambda\mathcal{B}$ by a supermatrix $\sigma$ with independent matrix elements. For the required power of superdeterminant in the expression for the generating function, we write a super-Fourier representation
\begin{equation}
\mbox{sdet}^{-T/2}(\mathbb{1}_4+i\mathcal{B}^t\Lambda\mathcal{B}) = \int d[\rho] I(\rho) \exp(- \displaystyle\frac{i}{2} \mbox{str}(\mathcal{B}^t\Lambda\mathcal{B}_\rho)) \;.
\label{eq-9}
\end{equation}
The Fourier transform gives a supersymmetric Ingham-Siegel distribution,
\begin{equation}
I(\rho) = \int d[\sigma] \mbox{sdet}^{-T/2}(\mathbb{1}_4+i\sigma)\exp(\displaystyle\frac{i}{2}  \mbox{str}(\sigma\rho)) \;. 
\label{eq-10}
\end{equation}
Here, $$\sigma = \begin{bmatrix} \sigma_0 & \tau \\ {\tau}^* & i\sigma_1 \mathbb{1}_2 \end{bmatrix} \mbox{and} \;\; \rho = \begin{bmatrix} \rho_0 & \omega \\ {\omega}^* & i\rho_1 \mathbb{1}_2 \end{bmatrix}$$ are supermatrices of dimension $4\times 4$ with real-symmetric $2\times2$ diagonal blocks. The off-diagonal blocks are Grassmann variables with the structure $$\tau = \begin{bmatrix} \eta & \eta^* \\ \xi & \xi^* \end{bmatrix} \mbox{and} \;\; \tau^* = \begin{bmatrix} \eta^* & \xi^* \\ -\eta & -\xi \end{bmatrix}$$ (similarly, for $\omega$). The super-integration measure is $d[\sigma] = (2\pi)^{-2} d\sigma_{0aa} d\sigma_{0ab} d\sigma_{0bb} d\sigma_{1} \partial\eta \partial\eta^* \partial\xi \partial\xi^*$, where $\sigma_{0aa}, \sigma_{0bb}$ are diagonal and $\sigma_{0ab}$ is the off-diagonal elements of $\sigma_0$. The measure $d[\rho]$ is defined in a similar fashion. Using these and integrating over the supermatrix $\mathcal{B}$, the generating function is a supermatrix integral,
\begin{equation}
Z(J) = \int d[\rho] I(\rho) \prod_{i=1}^N \mbox{sdet}^{-1/2}(x^+ \mathbb{1}_4 -J \gamma -\frac{1}{2}\rho \Lambda_i) \;.
\label{eq-11}
\end{equation}
Here, the matrix $\gamma = \mbox{diag}(0_2,-\mathbb{1}_2)$ is diagonal. For arbitrary small $J$, using Equations \eqref{eq-10} and \eqref{eq-11}, we have $Z(J) = \int d[\sigma] \int d[\rho] e^{\cal L}$ with a Lagrangian ${\cal L}$ given by
\begin{equation}
{\cal L} = - \displaystyle\frac{1}{2} \displaystyle\sum_{i=1}^N \mbox{str} \ln (x^+ \mathbb{1}_4 -\frac{1}{2}\rho \Lambda_i) + \displaystyle\frac{T}{2} \mbox{str} \ln \rho - \mbox{str} \rho \;,
\label{eq-12}
\end{equation}
and we end up with a scalar polynomial equation resulting from the saddle point equation that can be solved numerically,
\begin{equation}
 \displaystyle\frac{1}{2} \displaystyle\sum_{i=1}^N \displaystyle\frac{\Lambda_i}{2x^+-\rho_0 \Lambda_i} +
 \displaystyle\frac{T}{2\rho_0} -1 = 0 \;.
\label{eq-13}
\end{equation}
This is the main analytic result of the paper which we test with numerics for different WOE models in the following section. The one-point function is then given in terms of the complex solution, say $\rho_0({\bf x})$, of this saddle point equation, 
\begin{equation}
S({\bf x}) = -{2}\Im \rho_0({\bf x})/{\pi N {\bf x}} \;,
\label{eq-14}
\end{equation}
in the limit $N,T \to \infty$ with fixed ratio $\kappa = N/T$. Note that the eigenvalue density is normalized to unity. Equation \eqref{eq-13} is valid for CWOE with arbitrary correlations and the structure of the correlation matrix enters via its eigenvalues $\Lambda_i$ ($1 \leq i \leq N$). Equation \eqref{eq-13} is another version of a classical result \cite{Ma-67, Si-95, Ba-98, Wi-17}.

\section{Numerical results}

For the random selections, we have two choices: (a) 'Non-Singular Random Selection Ensemble' (NSRSE) in which a given time series can appear many times but at most once in the construction of any correlation matrix to avoid singularities. As mentioned above, we will have binomially many choices but the members of the ensemble are not entirely independent. We will usually not have $N$ and $T$ very large, but even so we will find that the behavior of the bulk is not significantly affected although the outliers are. Alternatively, for sufficiently large $N$ and $T$, we could use a random matrix model 'Exclusive Random Selection Ensemble' (ERSE) that constructs an ensemble that excludes any repetition of time series in its construction. We can use this ensemble to calculate the expectation values of the quantities we are interested in and average those over all or a subset of possible selections. In this case, we expect to a large extent coincidence with correlated Wishart ensembles but the procedure is rather complicated and we will thus focus on the first choice namely NSRSE.

We now proceed to analyze two special cases. First, we consider the case of constant correlations where we, in addition to the spectrum bulk, have an outlier that should be described. Here, we also consider the case of zero correlations i.e. uncorrelated time series, where we reproduce the Mar\u{c}enko-Pastur distribution \cite{Ma-67, Pa-72}. Then, we proceed to the block structure which we illustrate by using two blocks of time series which have constant internal correlation and relatively small correlation between the two blocks. Note that our results for NSRSE need not agree with theory for Wishart matrices because starting with a single representative of this ensemble in the large space, we select the smaller matrices from that space and repetitions of the selection will turn out to be important. We compare the distribution of outliers for NSRSE with ERSE in terms of the first four moments.

\subsection{Correlated Non-Singular Random Selection Ensemble With Constant Linear Correlations}

\begin{figure}
\centering
\includegraphics[width=6in]{fig1.eps}
\caption{Density of non-zero eigenvalues of a singular correlation matrix $C$ obtained from a data matrix $A$ of dimension $1000 \times 100$; $\kappa = 10$ with (a) micro-canonical and (c) canonical normalization. Ensemble averaged eigenvalue density for a 5000 member NSRSE of correlation matrices constructed using $0.9T \times T$ ($\kappa =0.9$) dimensional $A$ matrices with (b) micro-canonical and (d) canonical normalization. Numerical results are histograms and solid curves are obtained from Equation \eqref{eq-14}.}
\label{den-uncorr}	
\end{figure}

\begin{figure*}
\centering
\includegraphics[width=4in]{fig2.eps}
\caption{Ensemble averaged eigenvalue density for a 5000 member ensemble of $90 \times 90$ dimensional correlated NSRSE matrices with constant linear correlations defined by (a) $\upsilon=0.1$, (b) $\upsilon=0.5$ and (c) $\upsilon=0.9$. Here, $\kappa = 10$. Numerical results are histograms and solid curves are obtained from Equation \eqref{eq-13}. The solid histograms correspond to microcanonical normalization and empty histograms correspond to canonical normalization. Insets give the distribution of the outlier.}
\label{den-corr}	
\end{figure*}

\begin{table}[ht]
\caption{Moments for outliers with constant correlations. All values are listed as (NSRSE/ERSE).}
\centering
\begin{tabular}{c c c c c c}
\hline\hline
Case &  & Mean & Width & Skewness & Kurtosis \\ 
\hline\hline
Largest eigenvalue,  & $\upsilon=0.1$ & (10.69/11.94) & (1.23/0.61) & (0.21/0.05) & (-0.12/0.05) \\
correlated NSRSE/ERSE & $\upsilon=0.5$ & (45.53/48.90) & (3.18/0.93) & (-0.10/0.01) & (-0.12/0.12) \\
& $\upsilon=0.9$ & (80.97/82.19) & (1.17/0.53) & (-0.44/-0.01) & (0.32/0.13) \\
\hline
Largest eigenvalue, & $\upsilon_1=0.1$, $\upsilon_2=0.1$ & (8.96/10.27) & (1.03/0.53) & (0.25/0.08) & (0.03/-0.03) \\
block NSRSE/ERSE & $\upsilon_1=0.1$, $\upsilon_2=0.5$ & (36.66/41.07) & (3.09/0.79) & (-0.04/0.03) & (-0.07/-0.04) \\
& $\upsilon_1=0.5$, $\upsilon_2=0.1$ & (10.83/10.85) & (1.05/0.45) & (0.31/0.16) & (0.13/-0.03) \\
& $\upsilon_1=0.5$, $\upsilon_2=0.5$ & (36.71/40.92) & (3.12/0.79) & (-0.04/0.03) & (-0.04/-0.03) \\
\hline
Second largest eigenvalue, & $\upsilon_1=0.1$, $\upsilon_2=0.1$ & (3.87/3.87) & (0.32/0.28) & (0.47/0.30) & (0.16/-0.02) \\
block NSRSE/ERSE  & $\upsilon_1=0.1$, $\upsilon_2=0.5$ & (3.41/3.21) & (0.41/0.28) & (0.36/0.22) & (0.27/0.01) \\
& $\upsilon_1=0.5$, $\upsilon_2=0.1$ & (8.3/9.71) & (0.85/0.44) & (0.06/-0.04) & (0.04/-0.06) \\
& $\upsilon_1=0.5$, $\upsilon_2=0.5$ & (9.91/9.48) & (1.31/0.42) & (0.29/0.08) & (0.15/0.003) \\
\hline\hline
\end{tabular}
\label{table1}
\end{table}

We consider correlated NSRSE ${\cal A} = \chi^{1/2} A$ with constant linear correlations defined by $\chi_{j,k} = \delta_{j,k} + \upsilon \; (1-\delta_{j,k})$; $\upsilon$ being the correlation coefficient. NSRSE of correlation matrices will then be obtained from the data set ${\cal A}$ of correlated white noise time series by selecting $m$ time series in $L$ samples from the $\binom{N}{m}$ possible selections. The corresponding eigenvalues will be obtained numerically below and compared to the solution of the polynomial equation in Equation \eqref{eq-13}. The parameters used in the calculations are $L = 5000$, $N = 1000$, $\kappa = 10$ and $a = 0.9$. We choose constant linear correlations defined by $\upsilon=0$, $0.1$, $0.5$ and $0.9$.  For Monte-Carlo simulations, we start with a singular data matrix of dimension $1000 \times 100$ ($\kappa = 10$). One can normalize these $1000$ time series in two ways: (1) by rescaling each time series by its respective mean and standard deviation ({\it micro-canonical} normalization) and (2) by rescaling all the time series by their average mean and average standard deviation ({\it canonical} normalization). Then, by randomly selecting the rows of this data matrix as explained above, we construct a 5000 member ensemble of $90 \times 100$ ($a = 0.9$) data matrices ($\kappa =0.9$). Using these, we construct the $L=5000$ members of NSRSE and diagonalize these to obtain the eigenvalues. 

The choice $c=0$ corresponds to uncorrelated NSRSE and average eigenvalues $\Lambda_i =1$ for $i=1,\ldots,N$. Using these in Equation \eqref{eq-13} results in a quadratic equation which can be solved analytically to obtain
\begin{equation}
\rho_0(y) = \displaystyle\frac{\kappa}{2\pi y} \left[ (y_+ - y) (y - y_-) \right]^{1/2} \;.
\label{eq-14}
\end{equation} 
Here, $\kappa = {N}/{T}$ and $y_\pm = T\left( 1 \pm \sqrt{\kappa} \right)/4$ define the spectral support of the eigenvalue density. This describes the distribution of non-zero eigenvalues for WOE in the limit $N,T \to \infty$ with fixed $\kappa$.  Hence, in order to be able to compare with the Mar\u{c}enko-Pastur distribution and numerics, one needs to re-scale the variables as $x^+ \to x^\prime T/4$ and $\rho(x^+) \to 4\rho(x^\prime)/T$ in Equation \eqref{eq-13}.  We compare numerical NSRSE eigenvalue densities with the analytical result given by Equation \eqref{eq-14} in Figure \ref{den-uncorr}. In Figure \ref{den-uncorr}(a), we show the numerical histogram for the $1000$ eigenvalues of the correlation matrix corresponding to the initial $1000 \times 100$ data matrix obtained using microcanonical normalization and similarly for canonical normalization in Figure \ref{den-uncorr}(b). The spectral bounds are in agreement with the solid curve obtained using Equation \eqref{eq-14}. However, as we have a single copy of correlation matrix, there are a lot of fluctuations in numerics. We do not find any significant differences between microcanonical and canonical normalizations for NSRSE. Then, we apply ensemble technique and eigenvalue histograms for microcanonical and canonical normalizations respectively are shown in Figure \ref{den-uncorr}(c) and \ref{den-uncorr}(d). The agreement with the solid curves obtained using Equation \eqref{eq-14} is excellent. Again, we do not observe any significant differences in the microcanonical and canonical normalizations for NSRSE using the ensemble technique.

The numerical histograms obtained for correlated NSRSE with constant linear correlations defined by $\upsilon=0.1$, $0.5$ and $0.9$ are shown respectively in Figures \ref{den-corr}(a), \ref{den-corr}(b) and \ref{den-corr}(c). The solid histograms correspond to microcanonical normalization and empty histograms correspond to canonical normalization. The solid curves are obtained by numerically solving Equation \eqref{eq-13} with $\Lambda_i = 1-\upsilon$ for $i=1,\ldots,N-1$ and $\Lambda_N=N\upsilon+1-\upsilon$ (a third order polynomial equation). Insets in each of these pictures show the distribution of the outlier $\Lambda_N$. The agreement of the polynomial equation solution in the bulk of the spectrum is excellent except for small deviations in the tails with increasing correlation coefficient $\upsilon$. Notice the increasing difference between the bulk and the outlier along with shrinking of spectral bounds for the bulk distribution with increasing $\upsilon$. The histograms for microcanonical and canonical normalizations are similar for the bulk distribution while there are differences in outliers noticeable with increasing $\upsilon$. 
    
\begin{figure*}[!ht]
\centering
\includegraphics[width=6.5in]{fig3.eps}
\caption{Probability distribution of the outliers for NSRSE and ERSE. The largest eigenvalues are normalized with respect to their centroids ($\mu$) and widths ($\sigma$) i.e. $\hat{E} = (E_{max}-\mu)/\sigma$. The solid histograms correspond to NSRSE and the empty histograms correspond to ERSE. Corresponding first four moments are given in Table \ref{table1}.}
\label{corr-maxeig}	
\end{figure*}

The shape of the farthest peak (outlier) is Gaussian for the numerical histograms whereas it resembles a semicircle for the respective solutions from the polynomial equation. The saddle point approximation must be good where many peaks overlap. It must be worse for individual peaks (outliers). But as seen from Figure \ref{den-corr}, the saddle point approximation reproduces the position of the outliers not too far from reality. However, it cannot reproduce the shape of the peaks. In the saddle point approximation, the bulk of the spectrum is order $N$ correction and if the outlier is far away from the bulk, it is only order 1 correction term. The exact problem is highly complex and one cannot expect to get all the features by a simple polynomial equation.

It is now well established that the distribution of the largest eigenvalue separated from the bulk for a correlated covariance matrix converges to a Gaussian distribution \cite{Bai-04}. As ERSE should produce results close to Wishart ensembles, we compare the largest eigenvalue distributions for NSRSE and ERSE in Figure \ref{corr-maxeig}. The corresponding moments are given in Table \ref{table1}. As can be seen from these results, the largest eigenvalue distributions for NSRSE are also Gaussian, however the moments are different. Thus, the repetition of time series in the construction of NSRSE strongly affects the outliers.

\subsection{Block Non-Singular Random Selection Ensemble}

\begin{figure*}[ht]
\centering
\includegraphics[width=5in]{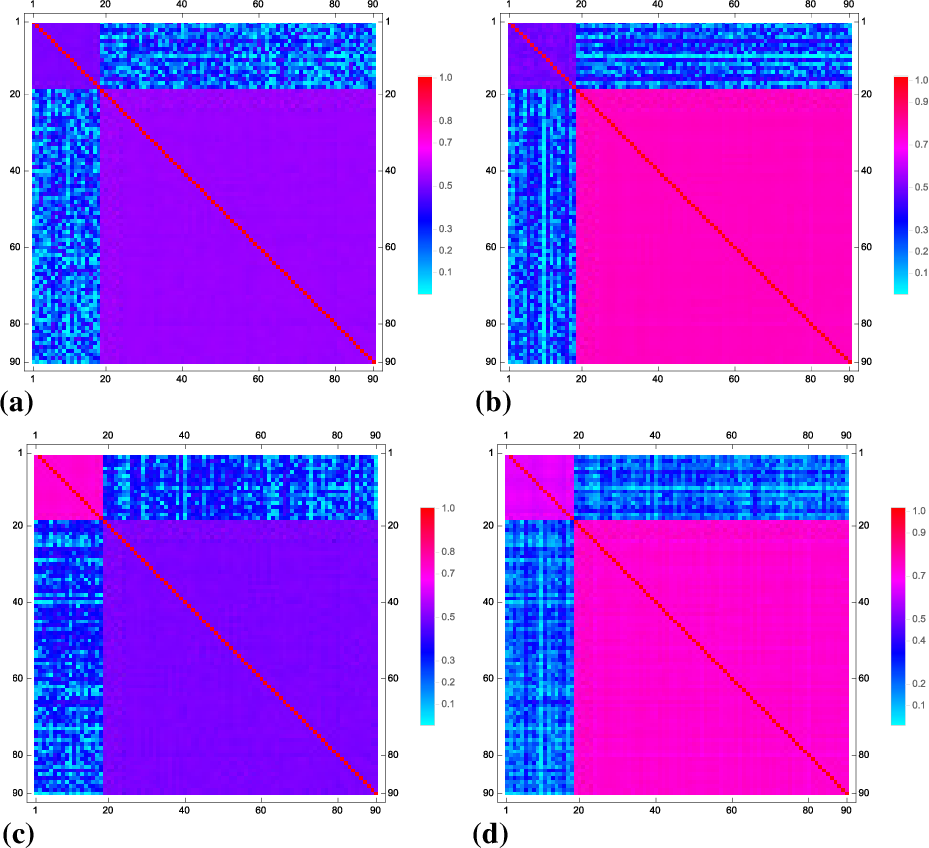}
\caption{Ensemble averaged non-singular block correlation matrices constructed using $90 \times 100$ dimensional data matrices $A$ with constant block correlations: (a) $\upsilon_1=\upsilon_2=0.1$; (b) $\upsilon_1=0.1$, $\upsilon_2=0.5$; (c) $\upsilon_1=0.5$, $\upsilon_2=0.1$ and (d) $\upsilon_1=\upsilon_2=0.5$.}
\label{matrix-stru}
\end{figure*}
    
\begin{figure*}[!ht]
\centering
\includegraphics[width=6.0in]{fig5.eps}
\caption{Ensemble averaged eigenvalue density for a 5000 member block NSRSE of correlation matrices constructed using $0.9T \times T$ ($\kappa =0.9$) dimensional $A$ matrices with constant block correlations defined by (a) $\upsilon_1=\upsilon_2=0.1$; (b) $\upsilon_1=0.1$, $\upsilon_2=0.5$; (c) $\upsilon_1=0.5$, $\upsilon_2=0.1$ and (d) $\upsilon_1=\upsilon_2=0.5$. Numerical results are histograms and solid curves are obtained from Equation \eqref{eq-13}. The solid histograms correspond to microcanonical normalization and empty histograms correspond to canonical normalization. Insets give the distribution of the outliers.}
\label{den-block}	
\end{figure*}

As is usual in financial market analysis, one deals with approximate block matrices where each block represents a sector. For instance, energy, utility and technology are a few sectors in stocks. Inspired by this, we consider a simple $2 \times 2$ model for a two sector NSRSE model. The corresponding data matrix has the structure $A^t = \left[ \begin{matrix} A_1 \; A_2 \end{matrix} \right]$ with $A_1$ and $A_2$ representing data matrix in each sector with respective dimensions $N_1 \times T$ and $N_2 \times T$; $N=N_1+N_2$. In each sector, we consider constant linear correlations with correlation coefficients $\upsilon_1$ and $\upsilon_2$.

For numerics, we construct a $L=5000$ member block NSRSE with $N =1000$, $\kappa =10$, $a = 0.9$, $N_1 = 0.2 N$ and $N_2 = 0.8 N$. To generate the ensemble, for each member, random selections of time series from the given $A$ matrix are done depending on the weights (say, these are $p_1$ and $p_2$): $p_1 =N_1*m/N$ and $p_2 =N_2*m/N$. These are the number of time series randomly chosen from each sector. 

One can also make the random permutations without any weights. This does not affect the structure of the correlation matrices in NSRSE. Figure \ref{matrix-stru} shows the structure of ensemble averaged correlation matrices constructed using canonical normalization with (a) $\upsilon_1=\upsilon_2=0.1$; (b) $\upsilon_1=0.1$, $\upsilon_2=0.5$; (c) $\upsilon_1=0.5$, $\upsilon_2=0.1$ and (d) $\upsilon_1=\upsilon_2=0.5$. Here random permutations were carried out without any weights. Thus, the block structure remains intact even without the weighted random permutations. This is obvious as the $\chi$ matrix is invariant under permutations.

In Figure \ref{den-block}, we compare the eigenvalue histograms (solid ones corresponding to microcanonical normalization and empty ones corresponding to canonical normalization) of block NSRSE for (a) $\upsilon_1=\upsilon_2=0.1$; (b) $\upsilon_1=0.1$, $\upsilon_2=0.5$; (c) $\upsilon_1=0.5$, $\upsilon_2=0.1$, (d) $\upsilon_1=\upsilon_2=0.5$ with the solid curve obtained using Equation \eqref{eq-13} with $\Lambda_i=1-\upsilon_1$ for $i=1,\ldots,N_1-1$, $\Lambda_{N_1}=N_1\upsilon_1+1-\upsilon_1$,  $\Lambda_i=1-\upsilon_2$ for $i=N_1+1,\ldots,N_2-1$, $\Lambda_{N_2}=N_2\upsilon_2+1-\upsilon_2$ (fifth order polynomial equation). We find good agreement in the bulk distributions with deviations in the tails for larger correlation coefficients $\upsilon_1$ and $\upsilon_2$. Insets show the distributions of the two outliers ($\Lambda_{N_1}$ and $\Lambda_{N_2}$). It can be single peaked, overlapping peaks or double peaked as the positions depend on correlation coefficients $\upsilon_1$ and $\upsilon_2$. The choice of normalization generates differences in the distribution of outliers. The saddle point approximation gives the approximate positions of the peaks but not the shape. 

We compare the distributions of the outliers (largest and second largest eigenvalues) for NSRSE with those corresponding to ERSE in Fig. \ref{corr-maxeig-block}. The corresponding moments are given in Table \ref{table1}. The distributions of outliers separated from the bulk are well approximated by Gaussians for both NSRSE and ERSE while the moments are different. The convergence to Gaussian distribution also depends on the separation of the outliers from the bulk distribution. Thus, the repetition of time series in the construction of block NSRSE strongly affects the outliers.

\begin{figure*}[!ht]
\centering
\includegraphics[width=6.5in]{fig6.eps}
\caption{Probability distribution of the outliers for NSRSE and ERSE. Upper panel [(a)-(d)] gives the distribution of largest eigenvalue for block correlation matrices constructed using $90 \times 100$ dimensional data matrices $A$ with constant block correlations: (a) $\upsilon_1=\upsilon_2=0.1$; (b) $\upsilon_1=0.1$, $\upsilon_2=0.5$; (c) $\upsilon_1=0.5$, $\upsilon_2=0.1$ and (d) $\upsilon_1=\upsilon_2=0.5$. Similarly, the lower panel [(e)-(h)] gives the distribution of second largest eigenvalue. In the upper panel, the largest eigenvalues are normalized with respect to their centroids ($\mu$) and widths ($\sigma$) i.e. $\hat{E} = (E_{max}-\mu)/\sigma$. Similarly, $\hat{E} = ((E_{max}-1)-\mu)/\sigma$ with the corresponding $\mu$ and $\sigma$ in the lower panel. The solid histograms correspond to NSRSE and the empty histograms correspond to ERSE. Corresponding first four moments are given in Table \ref{table1}.}
\label{corr-maxeig-block}	
\end{figure*}

\section{Conclusions and Outlook}

We have presented an entirely new way to treat large numbers of short time series pertaining to the same system and therefore likely to display some correlation. Basically the proposition consists in dividing the entire set of time series in different ways, thus obtaining the Non-Singular Random Selection Ensemble from the data. This allows to obtain a spectral distribution for an ensemble of correlation or covariance matrices and also to get distributions of particular eigenvalues, importantly the largest or the smallest one. Using our technique, we obtain a large set of eigenvalues for a singular data matrix and thus, get information about the bulk eigenvalue distribution along with the outliers, which is otherwise not possible. It also allows analysis of two and three point functions which is impossible for a small set (total number $T$ with $T << N$) of non-zero eigenvalues for a single data matrix. The same will hold for eigenfunctions. If the matrix elements are not Gaussian distributed, the shapes and moments of distributions of eigenvalues will change with the strength of the correlation coefficient and there will be effects in the tails of the distributions. We expect that our technique will be sensitive to tails of distributions of eigenvalues as the probability of randomly selecting the outliers increases with our technique. Finally we may note that selection of such subsets gives valuable insight into the consequences of having incomplete sets of time series. 
This is illustrated in Ising model calculations \cite{Pr-14} where random selections of subsets of very long time series were used to test the effect of having incomplete measurements, which showed that a power law remaines visible using only 10\% of the time series associated with the lattice. This opens a new alley for investigations of systems that are not stationary on longer time scales but quasi-stationary on a short time scale as defined by the length of the epochs we choose. We can then study the temporal evolution of such an ensemble. This in turn may give hints to instabilities emerging in the system which might be sufficiently strong to be used to give an early warning. 

The next step will be to show how such an ensemble behaves, when at or near a critical transition. At this point we are studying this in financial markets and in two dynamical systems, namely the TASEP \cite{Bi-17} and the 2-D Ising model near criticality \cite{Vi-13}. The range of potential applications is very wide and in the present paper we have performed the first tests using correlated random matrices as a model where analytic results are available. The case we generically discuss is a set of time series, which are strongly correlated within each of two subsets leading to a block structure in the correlation matrix. This is a toy model for financial markets with its traditional division into market sectors. Preliminary results on financial markets can be viewed in a master thesis \cite{JT} and further work in this direction is in progress.

\section*{Acknowledgements}

We thank F. Leyvraz and M. Kieburg for useful discussions. Authors acknowledge financial support from UNAM/DGAPA/PAPIIT research grant IA104617 and CONACyT FRONTERAS 201.

\section*{Author contributions statement}

Manan Vyas, Thomas Guhr and Thomas H. Seligman contributed to the entire development and writing of the paper in all aspects except for numerics carried out exclusively by Manan Vyas.

\section*{Additional information}

The authors declare no competing interests.

\section*{Data availability statement}

The datasets generated during and/or analyzed during the current study are available from the corresponding author on reasonable request.

\newpage

\end{document}